\documentclass[aps,preprint,showpacs,preprintnumbers,amsmath,amssymb]{revtex4}
\usepackage{amsmath,mathrsfs,amsbsy,color,graphicx,bm,amsthm,amsfonts}
\usepackage{units}
\usepackage{bbm}
\usepackage{times}
\usepackage{dcolumn}
\usepackage{mathrsfs}
\usepackage{amsmath,amssymb,epsfig}
\usepackage{amsmath}
\newcommand{\udots}{\mathinner{\mskip1mu\raise1pt\vbox{\kern7pt\hbox{.}}
\mskip2mu\raise4pt\hbox{.}\mskip2mu\raise7pt\hbox{.}\mskip1mu}}
\begin{document}
\title{Can boundary configuration be tuned to optimize directional quantum steering harvesting? }
\author{Xiao-Li Huang$^1$\footnote{ huangxiaoli1982@foxmail.com }, Xiao-Ying  Jiang$^1$, Yu-Xuan Wang$^1$, Si-Yu Liu$^1$,   Zejun Wang$^2$ \footnote{zejunwangcz@foxmail.com}, Shu-Min Wu$^1$\footnote{smwu@lnnu.edu.cn  (corresponding author)}}
\affiliation{$^1$  Department of Physics, Liaoning Normal University, Dalian 116029, China \\
$^2$  Department of Physics, Changzhi University, Changzhi, 046011, China
}


\begin{abstract}
We investigate the harvesting of quantum steering and its asymmetry between two static detectors locally interacting with a vacuum massless scalar field near an infinite, perfectly reflecting boundary. The detectors are arranged either parallel or orthogonal to the boundary, with detector $B$ assumed to have an energy gap greater than or equal to that of detector $A$.
It is interesting to observe that, with increasing distance between the detectors and the boundary, the boundary  tends to suppress quantum steering in one direction while enhancing it in the opposite direction. In the case of identical detectors, steering is symmetric when they are aligned parallel to the boundary. However, orthogonal alignment breaks this symmetry due to their unequal spatial proximity to the boundary. For non-identical detectors in the parallel configuration, the steering from $A$ to $B$ ($A \rightarrow B$) generally surpasses that from $B$ to $A$ ($B \rightarrow A$). In contrast, when the detectors are oriented orthogonally to the boundary, the relative strength of $A \rightarrow B$ and $B \rightarrow A$ steerability depends on the interplay between the boundary effects and the detectors' energy gap difference. Across most of the parameter space, the orthogonal alignment tends to enhance $B \rightarrow A$ steering while suppressing $A \rightarrow B$ steering compared to the parallel setup. These findings suggest that boundary configurations should be flexibly adjusted according to the directional dependence of steering harvesting in order to optimize quantum information extraction.
\end{abstract}

\vspace*{0.5cm}
\maketitle
\section{Introduction}
The Einstein-Podolsky-Rosen (EPR) steering phenomenon is a fundamental manifestation of quantum nonlocality, wherein one party, typically Alice, can nonlocally influence the state of another party, Bob, through local measurements on an entangled system  \cite{A1,A2,A3,A4}. First introduced by  Schr\"{o}dinger in response to the original EPR paradox \cite{A6,A7,A8}, quantum steering has since emerged as a distinct class of quantum correlation-stronger than entanglement but generally weaker than Bell nonlocality \cite{A17}. One of the most intriguing and defining characteristics of EPR steering is its inherent directional asymmetry: it is possible for Alice to steer Bob's state, while the converse-Bob steering Alice may not be achievable. This asymmetry is not merely theoretical but has been robustly confirmed through experiments in both discrete-variable and continuous-variable quantum systems \cite{QGL1,wd1,wd2,wd3,wd4,wd5,wd6,tyR4,QGL2}. The directionality makes steering especially valuable in scenarios where trust is not symmetric between parties, enabling one-sided device-independent quantum protocols. In recent years, EPR steering has attracted growing attention not only for its foundational implications in the study of quantum correlations, but also for its practical utility. It plays a pivotal role in quantum cryptographic tasks such as quantum key distribution and quantum secret sharing, particularly under asymmetric trust settings \cite{A2,A3}. Moreover, it has found promising applications in quantum metrology, where steerable states can enhance measurement precision, and in quantum networks, where asymmetric correlations can be exploited for efficient information transfer \cite{hxl1,hxl2,hxl3,hxl4,hxl5}. These diverse applications underscore the versatility and significance of quantum steering as a resource in modern quantum information science.

The study of relativistic quantum information is deeply intertwined with the properties of the quantum vacuum  \cite{E27,zzzRM8,QGL3,QGL4,QGL6,QGL7,QGL8,QGL9,QGL10,QGL11,QGL12,QGL13,QGL14,QGL15}, whose state is inherently shaped by the underlying spacetime geometry  \cite{E26}. Prior research has revealed that the vacuum of quantum fields exhibits intrinsic nonlocal correlations \cite{L15,L17,L18,L19,L40}, which can be extracted using Unruh-DeWitt (UDW) detectors. These harvested correlations manifest as quantum entanglement, discord, and mutual information between the detectors, and can even be distilled into Bell pairs, thereby establishing the quantum vacuum as a valuable resource for quantum information processing. The process, referred to as entanglement harvesting, has attracted significant attention, particularly under nontrivial physical conditions, such as the presence of reflecting boundaries, curved spacetime backgrounds, or noninertial detector motion \cite{L36,L37,L38,L39,L41,L42,L42-3,L42-4,L42-5,L42-7,L42-8,L42-9,L42-10,L42-11,L42-12,L42-13,L42-15,L26,L27,L28,L29,L30,L31,L32,QTL32}. A particularly striking result is that, even when the detectors are placed asymmetrically with respect to a boundary, the harvested entanglement remains symmetric and exhibits no directional bias  \cite{L42-11,L26}. This observation suggests that quantum entanglement, while a powerful indicator of nonclassicality, may not fully capture subtle spatial asymmetries embedded in boundary-influenced field configurations. In contrast to entanglement, quantum steering exhibits a richer structure, characterized by its inherent directional asymmetry. This feature gives rise to a hierarchy of steerability: two-way steering, one-way steering, and no-way steering, each with distinct physical and operational implications. Motivated by this directionality, we investigate whether the introduction of a boundary, relative to the boundary-free case, tends to suppress steering from Bob to Alice while simultaneously enhancing it in the reverse direction.  We further probe whether the directional nature of quantum steering requires adaptive boundary configurations for optimal harvesting. Exploring these questions not only deepens our understanding of quantum steering in boundary-influenced quantum field theory, but also extends harvesting protocols to encompass directional  quantum correlations beyond entanglement.

In this paper, we present a comprehensive investigation of the directional harvesting of quantum steering by two UDW detectors, labeled $A$ and $B$, placed near a perfectly reflecting boundary. Without loss of generality, we assume that detector $B$ has an energy gap equal to or greater than that of detector $A$. Our primary objective is to explore how the presence of the boundary influences the steering harvested by the detectors, especially in scenarios where the detectors are aligned either parallel or orthogonal to the boundary and possess differing energy gaps. This study focuses on three central aspects: (i) how the boundary affects the asymmetry of the harvested quantum steering;  (ii) how the achievable separation distance and the magnitude of steering-specifically $A\rightarrow B$ and $B\rightarrow A$ vary under different alignment configurations; (iii)  how to appropriately tailor boundary configurations in response to the directionality of quantum steering. Our results reveal that, within certain regions of the parameter space, the boundary can suppress steering in one direction while simultaneously enhancing it in the opposite direction. We also find that boundary configurations must be adaptively chosen to efficiently harvest quantum steering depending on the desired direction. Moreover, the asymmetry in the detectors' energy gaps can significantly influence the amount of harvested steering. Optimal steering is achieved when the gap difference matches specific inter-detector separations and distances from the boundary.  Overall, in contrast to quantum entanglement, which typically remains symmetric under such conditions, the steering harvested from the vacuum quantum field exhibits pronounced directionality.

The structure of this paper is as follows. Sec. II presents a brief overview of the quantification of quantum steering. In Sec. III, within the framework of the UDW detector model, we derive analytical expressions for vacuum quantum steering in both directions within the framework of the UDW detector model. Sec. IV investigates in detail the influence of the reflective boundary and the detectors' energy gaps on the harvested steering, accompanied by a comparative analysis of the results for the parallel and orthogonal configurations. For simplicity, we adopt natural units throughout the paper by setting $\hbar = c = k_B = 1$.

\section{Quantification of quantum steering}
It is well known that quantum steering is a type of quantum correlation between Bell nonlocality and quantum entanglement. Recently, it was discovered that quantum steering can be indirectly detected by quantum entanglement \cite{B1,B2}.
In this paper, we consider a special type of quantum state, the $X$-state, whose density matrix $\rho _{AB}^X$ shared by Alice and Bob can be written as
\begin{eqnarray}\label{S1}
\rho_{AB}^X=\left(\!\!\begin{array}{cccccccc}
\rho_{11} & 0 & 0 & \rho_{14}\\
0 & \rho_{22} & \rho_{23} & 0\\
0 & \rho_{32} & \rho_{33} & 0\\
\rho_{41} & 0 & 0 & \rho_{44}\\
\end{array}\!\!\right),
\end{eqnarray}
where the real elements of the matrix satisfy $\rho_{ij}=\rho_{ji}$.
Then, we use the concurrence to measure quantum entanglement \cite{B3}, and its analytical expression is shown in the following form \cite{B4}
\begin{eqnarray}\label{S2}
C(\rho_{AB}^X)=2\max\ \big\{0,|\rho_{14}|-\sqrt{\rho_{22}\rho_{33}}, |\rho_{23}|-\sqrt{\rho_{11}\rho_{44}}\ \big\}.
\end{eqnarray}

For a general bipartite state $\rho_{AB}$ shared by Alice and Bob, the steering from Bob to Alice can be observed if the density matrix $\tau_{AB}$ defined as \cite{B2,B6}
\begin{eqnarray}\label{S3}
\tau_{AB}=\frac{\rho_{AB}}{\sqrt{3}}+\frac{3-\sqrt{3}}{3}(\rho_{A}\otimes\frac{I}{2}),
\end{eqnarray}
is entangled, where $I$ represents the two-dimensional identity matrix and $\rho_{A}=Tr_{B}(\rho_{AB})$. Similarly, we can observe the corresponding steering from Alice to Bob if the state $\tau_{BA}$
defined as
\begin{eqnarray}\label{S4}
\tau_{BA}=\frac{\rho_{AB}}{\sqrt{3}}+\frac{3-\sqrt{3}}{3}(\frac{I}{2}\otimes\rho_{B}),
\end{eqnarray}
is entangled, where $\rho_{B}=Tr_{A}(\rho_{AB})$. Using Eqs.(\ref{S1}) and (\ref{S3}), through simple calculations, the density matrix $\tau^{X}_{AB}$ of Eq.(\ref{S1}) can be represented as
\begin{eqnarray}\label{S5}
\tau^{X}_{AB}=\left(\!\!\begin{array}{cccccccc}
\frac{\sqrt{3}}{3}\rho_{11}+m & 0 & 0 & \frac{\sqrt{3}}{3}\rho_{14}\\
0 & \frac{\sqrt{3}}{3}\rho_{22}+m & \frac{\sqrt{3}}{3}\rho_{23} & 0\\
0 & \frac{\sqrt{3}}{3}\rho_{32} & \frac{\sqrt{3}}{3}\rho_{33}+n & 0\\
\frac{\sqrt{3}}{3}\rho_{41} & 0 & 0 & \frac{\sqrt{3}}{3}\rho_{44}+n\\
\end{array}\!\!\right),
\end{eqnarray}
with $m=\frac{(3-\sqrt{3})}{6}(\rho_{11}+\rho_{22})$ and $n=\frac{(3-\sqrt{3})}{6}(\rho_{33}+\rho_{44})$. By using Eq.(\ref{S2}), if one of the following inequalities is satisfied, the state $\tau_{AB}^X$ is entangled,
\begin{eqnarray}\label{S6}
|\rho_{14}|>\sqrt{G_{a}-G_{b}},
\end{eqnarray}
or
\begin{eqnarray}\label{S7}
|\rho_{23}|>\sqrt{G_{c}-G_{b}},
\end{eqnarray}
 where
\begin{gather}\label{S8}
G_{a}=\frac{2-\sqrt{3}}{2}\rho_{11}\rho_{44}+\frac{2+\sqrt{3}}{2}\rho_{22}\rho_{33}+\frac{1}{4}(\rho_{11}+\rho_{44})(\rho_{22}+\rho_{33}),\nonumber\\
G_{b}=\frac{1}{4}(\rho_{11}-\rho_{44})(\rho_{22}-\rho_{33}),\nonumber\\
G_{c}=\frac{2+\sqrt{3}}{2}\rho_{11}\rho_{44}+\frac{2-\sqrt{3}}{2}\rho_{22}\rho_{33}+\frac{1}{4}(\rho_{11}+\rho_{44})(\rho_{22}+\rho_{33}).
\end{gather}
Employing a similar method, we can observe that the steering from Alice to Bob can be discerned through one of the following inequalities
\begin{eqnarray}\label{S9}
|\rho_{14}|>\sqrt{G_{a}+G_{b}},
\end{eqnarray}
or
\begin{eqnarray}\label{S10}
|\rho_{23}|>\sqrt{G_{c}+G_{b}}.
\end{eqnarray}
Based on the above inequalities, we can quantify the steering from Bob to Alice
\begin{eqnarray}\label{S11}
S^{B\rightarrow A}=\max\ \big\{0, |\rho_{14}|-\sqrt{G_{a}-G_{b}}, |\rho_{23}|-\sqrt{G_{c}-G_{b}}\ \big\},
\end{eqnarray}
and the steering  from Alice to Bob
\begin{eqnarray}\label{S12}
S^{A\rightarrow B}=\max\ \big\{0, |\rho_{14}|-\sqrt{G_{a}+G_{b}}, |\rho_{23}|-\sqrt{G_{c}+G_{b}}\ \big\}.
\end{eqnarray}

We notice that the $S^{B\rightarrow A}$ and the $S^{A\rightarrow B}$ are often different, which spark our research interest in exploring asymmetric quantum steering harvested from the quantum vacuum in the presence of a reflecting boundary.  Thus, quantum steering manifests in three distinct scenarios:  (i) no-way steering, indicating the state lacks steerability in any direction; (ii) two-way steering, where the state is steerable in both directions;  (iii) one-way steering, implying steerability in only one direction. This last scenario highlights the asymmetry inherent in quantum steering. To quantify this steering asymmetry, we define it between the detector $A$ and the detector $B$ as
\begin{eqnarray}\label{S13}
S^\Delta_{AB}= S^{A\rightarrow B}-S^{B\rightarrow A}.
\end{eqnarray}

\section{Vacuum steering in both directions for UDW detector model}
In the standard steering harvesting protocol, one considers two UDW detectors, denoted as $A$ and $B$, which interact locally with a massless quantum scalar field $\phi[x_D(\tau)] \quad  (D\in\{A,B\})$ along their respective worldlines. The detector's classical trajectory, $ x_D(\tau)$, is parameterized with respect to its proper time $\tau$. Then, the interaction Hamiltonian for a detector locally coupled to the scalar field in the interaction picture is denoted by
\begin{eqnarray}\label{s14}
H_{D}(\tau)=\lambda\chi(\tau)\big[e^{i\Omega_D\tau}\sigma^++e^{-i\Omega_D\tau}\sigma^-\big]\phi[x_D(\tau)], \quad  D\in\{A,B\},
\end{eqnarray}
where $\lambda$ is the coupling strength. Here, each detector is modeled as a two-level quantum system with an energy gap $\Omega_D$ between the ground state $|0\rangle_D$ and the excited state $|1\rangle_D$. These operators are the SU(2) ladder operators: $\sigma^+=|1\rangle_D\langle0|_D$ and $\sigma^-=|0\rangle_D\langle 1|_D$.
To regulate the temporal profile of the detector-field interaction, we introduce a Gaussian switching function $\chi(\tau)=\exp[-\tau^2/(2\sigma^2)]$, where the parameter
$\sigma$ characterizes the duration of the interaction. Notably, all relevant physical quantities in the model can, in principle, be rendered dimensionless by rescaling with respect to $\sigma$.

Assuming both detectors are in their ground state, $|0\rangle_A$ and $|0\rangle_B$, and the field is in the Minkowski vacuum state $|0\rangle_M$, then the initial state of the system is $|\Psi\rangle_i=|0\rangle_A|0\rangle_B|0\rangle_M$. Using the interaction Hamiltonian defined in Eq.(\ref{s14}), the time evolution of the system in the interaction picture is expressed as
\begin{eqnarray}\label{S15}
|\Psi\rangle_f:=\mathcal{T}\exp\bigg[-i\int dt\bigg({\frac{d\tau_A}{dt}}H_A(\tau_A)+{\frac{d\tau_B}{dt}}H_B(\tau_B)\bigg)\bigg]|\Psi\rangle_i,
\end{eqnarray}
where $\mathcal{T}$ is the time ordering operator and $t$ is the coordinate time associated with the field's vacuum reference frame.
By tracing out the field degrees of freedom and applying perturbation theory, one can derive that the density matrix for the final state of the detectors, in the basis
 $\{|0\rangle_A|0\rangle_B, |0\rangle_A|1\rangle_B, |1\rangle_A|0\rangle_B, |1\rangle_A|1\rangle_B\}$, at leading order in the coupling strength, is represented by \cite{L37,AH1}
\begin{eqnarray}\label{S16}
\rho_{AB} :=\rm{tr}_{\phi}(U|\Psi_{f}\rangle\langle\Psi_{f}|U^\dagger )
=
\begin{pmatrix}
1-P_A-P_B & 0 & 0 &X\\
0 & P_B & C & 0\\
0 & C^\ast & P_A & 0\\
X^\ast & 0 & 0 & 0\\
\end{pmatrix}
+ \mathcal{O}\!\left(\lambda^4\right),
\end{eqnarray}
where the excitation probability for each detector is given by
\begin{eqnarray}\label{S17}
P_D:=\lambda^2\int\int d\tau d\tau'\chi(\tau)\chi(\tau')e^{-i\Omega_D(\tau-\tau')}W(x_D(t),x_D(t')) , D\in \{A,B\},
\end{eqnarray}
and the correlation terms between the detectors are defined as
\begin{eqnarray}\label{S18}
C:=\lambda^2\int\int d\tau d\tau'\chi(\tau)\chi(\tau')e^{-i(\Omega_A\tau-\Omega_B\tau')}W(x_A(t),x_B(t')) ,
\end{eqnarray}

\begin{eqnarray}\label{S19}
   \nonumber X:= -\lambda^2 \int\int &d\tau d\tau'\chi(\tau)\chi(\tau')e^{-i(\Omega_A\tau+\Omega_B\tau')}\big[\theta (t'-t)W(x_A(t),x_B(t'))\\
   &+ \theta(t-t')W(x_B(t'),x_A(t))\big] .
\end{eqnarray}
Here, $\theta(t)$ is the Heaviside theta function, and $W(x,x'):=\langle0|_M\phi(x)\phi(x')|0\rangle_M$ is the Wightman function of the quantum field in the Minkowski vacuum state.
If the detectors are in a stationary state, it means that $t = \tau$, indicating that the detector's coordinate time and proper time are identical.
By combining Eqs.(\ref{S11}), (\ref{S12}), and (\ref{S16}), we can get the specific expressions of the steering from Bob to Alice
\begin{equation}\label{S20}
\begin{aligned}
S^{B\rightarrow A}=\max \bigg\{0, &|X|-\sqrt{\frac{1+\sqrt{3}}{2}P_A P_B+\frac{1}{2}P_A-\frac{1}{2}{P_A}^2},\\
&|C|-\sqrt{\frac{1-\sqrt{3}}{2}P_A P_B+\frac{1}{2}P_A-\frac{1}{2}{P_A}^2}\bigg\},
\end{aligned}
\end{equation}
and the steering from Alice to Bob
\begin{equation}\label{S21}
\begin{aligned}
S^{A\rightarrow B}=\max \bigg\{0, &|X|-\sqrt{\frac{1+\sqrt{3}}{2}P_A P_B+\frac{1}{2}P_B-\frac{1}{2}{P_B}^2},\\
&|C|-\sqrt{\frac{1-\sqrt{3}}{2}P_A P_B+\frac{1}{2}P_B-\frac{1}{2}{P_B}^2}\bigg\}.\
\end{aligned}
\end{equation}
To investigate whether the quantum steering $S^{A\rightarrow B}$ and $S^{B\rightarrow A}$ are symmetric, we also calculate their difference, which we define as the steering asymmetry $S^\Delta_{AB}= S^{A\rightarrow B}-S^{B\rightarrow A}$. In the following sections, we will explore quantum steering from Bob to Alice and from Alice to Bob between two static detectors positioned near an ideal reflective boundary.  A perfectly reflecting plane boundary is assumed to be at $\Delta z=0$.

\section{Quantum steering for detectors aligned parallel and orthogonal to the boundary}
In this section, we investigate the influence of the reflecting boundary on quantum steering between two static Unruh-DeWitt detectors. To offer clearer insight into the boundary-induced effects, we present schematic diagrams illustrating the spatial configurations of the detectors. Specifically, we consider two distinct alignments: one in which the detectors are positioned parallel to the boundary, and another where they are arranged orthogonally to it.

\subsection{Quantum steering for the detectors aligned parallel to the boundary}
\begin{figure}
\begin{minipage}[t]{0.5\linewidth}
\centering
\includegraphics[width=3.0in,height=5.2cm]{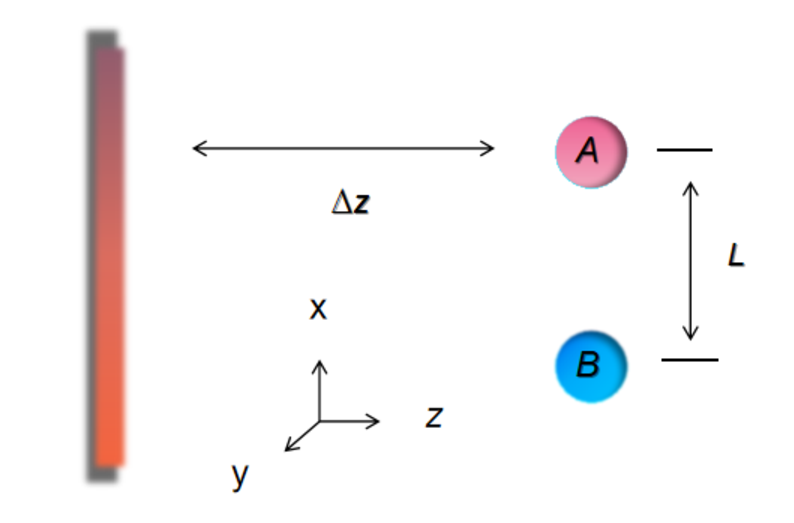}
\label{fig1a}
\end{minipage}%
\caption{ Schematic illustration of two static UDW detectors ($A$ and $B$) aligned parallel to a reflecting boundary. The detectors are separated by a distance $L$ along the vertical direction and are located at a fixed distance $\Delta z$.}
\label{Fig.1}
\end{figure}

In this subsection, we consider two static detectors separated by a distance $L$, with their distance from the boundary denoted by  $\Delta z$, and both detectors are aligned parallel to the boundary (see Fig.\ref{Fig.1}). Consequently, the spacetime trajectories of the detectors can be represented as
\begin{eqnarray}\label{S23}
x_A:=\{t_A,x=0,y=0,z=\Delta z\},\quad x_B:=\{t_B,x=L,y=0,z=\Delta z\}.
\end{eqnarray}

The transition probabilities $P_D$, as well as the correlation terms $C$ and $X$, can be computed using Eqs.(\ref{S17}), (\ref{S18}), and (\ref{S19}), respectively. To perform these calculations, one needs the Wightman function for vacuum massless free scalar fields, which can be derived using the method of images \cite{B9}
\begin{eqnarray}\label{S24}
 \nonumber W(x_D,x'_D)=-{\frac{1}{4\pi ^2}}\bigg[{\frac{1}{(t-t'-i\epsilon)^2-(x-x')^2-(y-y')^2-(z-z')^2}}\\
-{\frac{1}{(t-t'-i\epsilon)^2-(x-x')^2-(y-y')^2-(z+z')^2}}\bigg].
\end{eqnarray}
By substituting trajectory Eqs.(\ref{S23}) and (\ref{S24}) into Eq.(\ref{S17}), the transition probabilities can be obtained directly as
\begin{equation}\label{S25}
\begin{aligned}
 P_D=&{\frac{\lambda^2}{4\pi}}\bigg[e^{-\Omega_D ^2\sigma^2}-\sqrt{\pi}\Omega_D\sigma \rm{Erfc}(\Omega_D\sigma)\bigg]-{\frac{\lambda^2\sigma e^{-\Delta z^2/\sigma^2}}{8\sqrt{\pi }\Delta z}}\\
&\times\ \bigg\{\rm{Im}\bigg[e^{2i\Omega_D\Delta z}\rm{Erf}(i{\frac{\Delta z}{\sigma}}+\Omega_D\sigma)\bigg]-\sin(2\Omega_D \Delta z) \ \bigg\},
D\in \{A,B\},
\end{aligned}
\end{equation}
where $\rm{Erf(x)}$ is the error function and $\rm{Erfc(x)}:=1- \rm{Erf(x)}$ \cite{B10}. Similarly, the correlation terms $C$ and $X$ in this case can also be calculated as
\begin{eqnarray}\label{S26}
C={\frac{\lambda^2}{4\sqrt{\pi}}}e^{-{\frac{(\Omega_B-\Omega _A)^2\sigma^2}{4}}}\big[f(L)-f(\sqrt{L^2+4\Delta z^2)}\big],
\end{eqnarray}
\begin{eqnarray}\label{S27}
X=-{\frac{\lambda^2}{4\sqrt{\pi}}}e^{-{\frac{(\Omega_B+\Omega _A)^2\sigma^2}{4}}}\big[g(L)-g(\sqrt{L^2+4\Delta z^2)}\big],
\end{eqnarray}
with the auxiliary functions $f(L)$ and $g(L)$ defined as
\begin{eqnarray}\label{A1}
 f(L):={\frac{\sigma e^{-L^2/(4\sigma^2)}}{L}}\ \bigg\{\rm{Im}\bigg[e^{i(\Omega_A+\Omega_B)L/2}\rm{Erf}\bigg({\frac{iL+\Omega_B\sigma^2+\Omega_A\sigma^2}{2\sigma}}\bigg)\bigg]-\sin\bigg[{\frac{(\Omega_A+\Omega_B)L}{2}}\bigg]\bigg\},
\end{eqnarray}
\begin{eqnarray}\label{A2}
\nonumber g(L):=&{\frac{\sigma e^{-L^2/(4\sigma^2)}}{L}}\ \bigg\{\rm{Im}\bigg[e^{i(\Omega_B-\Omega_A)L/2}\rm{Erf}\bigg({\frac{iL+\Omega_B\sigma^2-\Omega_A\sigma^2}{2\sigma}}\bigg)\bigg]\\
 &+i\cos\bigg({\frac{(\Omega_B-\Omega_A)L}{2}}\bigg)\bigg\}.
\end{eqnarray}
Without loss of generality, we assume throughout this work that detector $B$ has an energy gap no smaller than that of detector $A$, i.e., $\Delta\Omega:=\Omega_B-\Omega_A\geq0$. It follows from Eqs.(\ref{S26}) and (\ref{S27}) that when the duration parameter is sufficiently small compared to the energy gap difference $(1/\sigma\ll\Delta\Omega)$, the correlation terms $C$ and $X$ tend to vanish.

\begin{figure}
\begin{minipage}[t]{0.5\linewidth}
\centering
\includegraphics[width=3.0in,height=5.2cm]{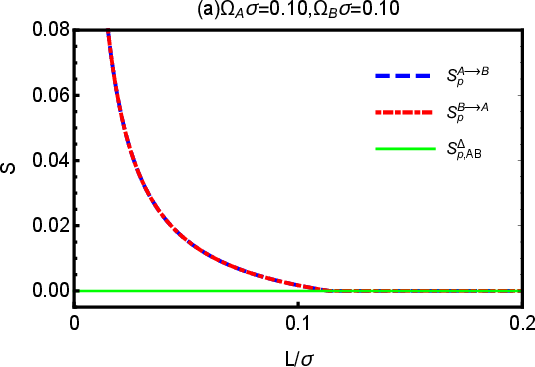}
\label{fig1a}
\end{minipage}%
\begin{minipage}[t]{0.5\linewidth}
\centering
\includegraphics[width=3.0in,height=5.2cm]{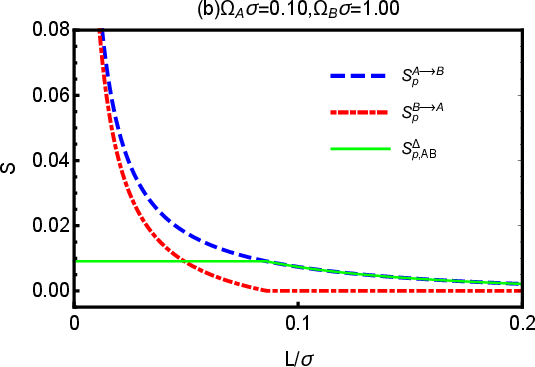}
\label{fig1b}
\end{minipage}%
\caption{Quantum steering $S^{A\rightarrow B}_P$, $S^{B\rightarrow A}_P$, and the steering asymmetry $S^\Delta_{P,{AB}}$ as a function of the detector separation $L/\sigma$ for various values of $\Omega_B\sigma$ with $\Delta z/\sigma=1.00$.}
\label{Fig.2}
\end{figure}

To gain a deeper understanding of the steerability of the two detectors in the presence of a reflective boundary, we employ numerical calculations. Initially, we illustrate the impact of a reflective boundary on the steerability of the two detectors aligned parallel to a reflecting boundary, where we plot the quantum steering $S^{A\rightarrow B}_P$, $S^{B\rightarrow A}_P$, and the steering asymmetry $S^\Delta_{P,{AB}}$ as a function of the detector separation $L/\sigma$ for different $\Omega_B\sigma$ in Fig.\ref{Fig.2}. Our observations reveal that as the separation of the detector $L/\sigma$ increases, the steerability in the case of identical detectors  decreases monotonically to zero, indicating a ``sudden death" of quantum steering with the $L/\sigma$. Furthermore, when the two detectors have different energy gaps, the steerability $S^{A\rightarrow B}_P$ is always bigger than the steerability $S^{B\rightarrow A}_P$, implying that the observer with the smaller energy gap experiences stronger steerability.  This demonstrates that steering harvesting is asymmetric due to the difference in energy gaps. The ``sudden death" of the quantum steering $S^{B\rightarrow A}_P$ suggests that the system $\rho_{AB}$ is undergoing a transition from two-way steering to one-way steering. Similarly, the ``sudden death" of the quantum steering $S^{A\rightarrow B}_P$ means that the system $\rho_{AB}$ is transitioning from one-way steering to no-way steering. Compared to the case of identical detectors, the gap difference enlarges the separation range over which steering from $A$ to $B$ ($S^{A\rightarrow B}_P$) can be harvested, while simultaneously suppressing the range for steering in the opposite direction ($S^{B\rightarrow A}_P$). These observations clearly demonstrate that quantum steering is not only direction-dependent but also highly sensitive to the intrinsic energy gap properties of the detectors.

\subsection{Quantum steering for detectors aligned orthogonally to the boundary}
\begin{figure}
\begin{minipage}[t]{0.5\linewidth}
\centering
\includegraphics[width=3.0in,height=5.2cm]{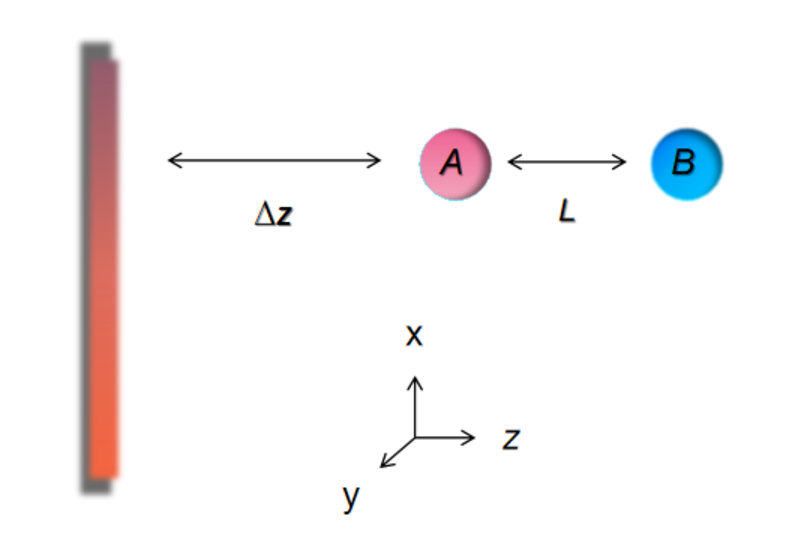}
\label{fig1a}
\end{minipage}%
\caption{Two static detectors are separated by a distance $L$, oriented orthogonally to the boundary plane. The distance $\Delta z$ represents the separation between the boundary and the detector that is positioned closer to it.}
\label{Fig.3}
\end{figure}

It is well known that the presence of a reflecting boundary in flat spacetime breaks spatial isotropy. As a result, the orientation of a detector pair relative to the boundary significantly influences the harvesting of  quantum steering. In this subsection, we analyze the scenario in which the two detectors are aligned orthogonally to the boundary plane. The spacetime trajectories of the two static detectors  [see Fig.\ref{Fig.3}] are given by
\begin{eqnarray}\label{S28}
x_A:=\{t_A,x=0,y=0,z=\Delta z\}, \quad x_B:=\{t_B,x=0,y=0,z=L+\Delta z\}.
\end{eqnarray}

\begin{figure}[htbp]
\centering
\includegraphics[height=1.8in,width=2.0in]{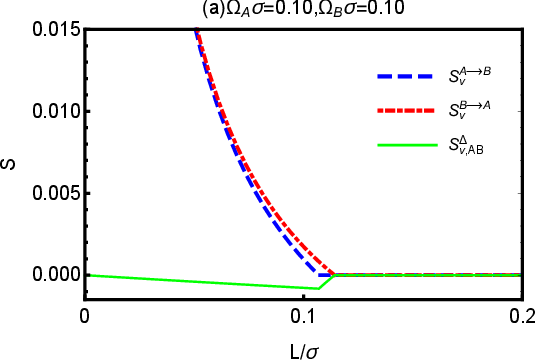}
\includegraphics[height=1.8in,width=2.0in]{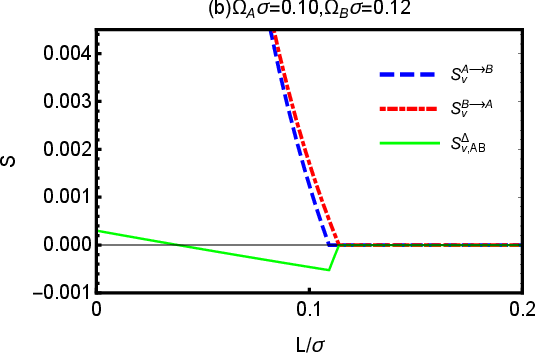}
\includegraphics[height=1.8in,width=2.0in]{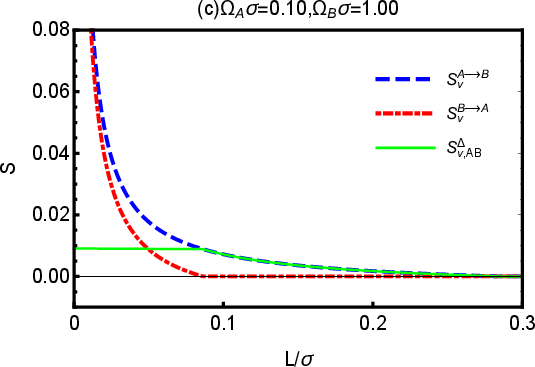}
\caption{Quantum steering $S^{A\rightarrow B}_V$, $S^{B\rightarrow A}_V$, and the steering asymmetry $S^\Delta_{V,{AB}}$ of two detectors as a function of $L/\sigma$ with $\Omega_A\sigma=0.10$ and $\Delta z/\sigma$=1.00.}\label{Fig.4}
\end{figure}

\begin{figure}
\begin{minipage}[t]{0.5\linewidth}
\centering
\includegraphics[width=3.0in,height=5.2cm]{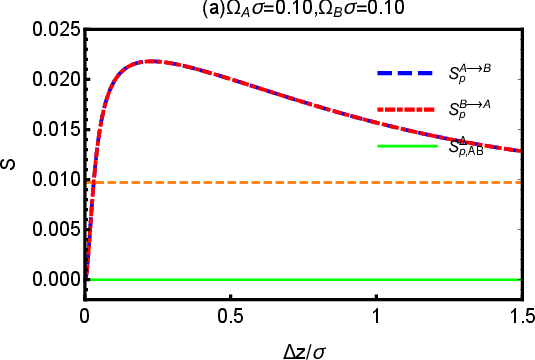}
\label{fig5a}
\end{minipage}%
\begin{minipage}[t]{0.5\linewidth}
\centering
\includegraphics[width=3.0in,height=5.2cm]{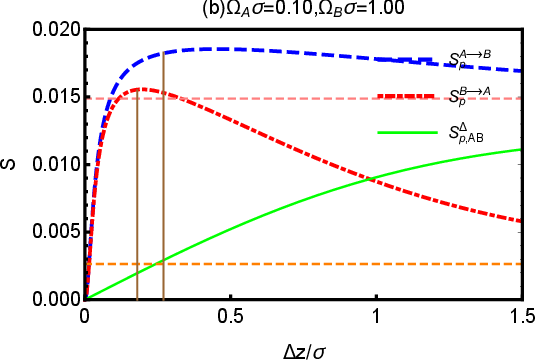}
\label{fig5b}
\end{minipage}%

\begin{minipage}[t]{0.5\linewidth}
\centering
\includegraphics[width=3.0in,height=5.2cm]{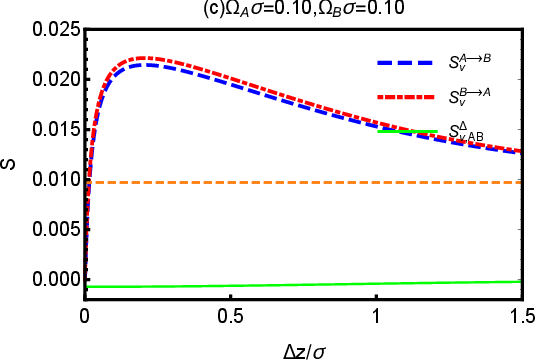}
\label{fig5c}
\end{minipage}%
\begin{minipage}[t]{0.5\linewidth}
\centering
\includegraphics[width=3.0in,height=5.2cm]{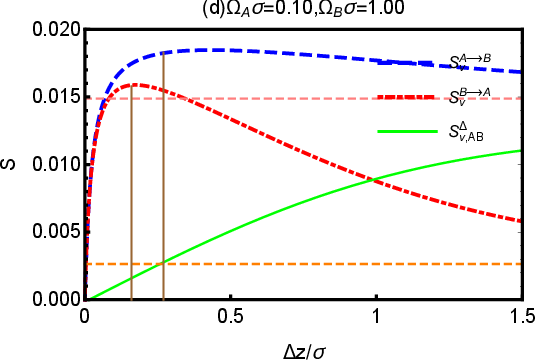}
\label{fig5d}
\end{minipage}%
\caption{The quantum steering $S^{A\rightarrow B}$, $S^{B\rightarrow A}$, and the steering asymmetry $S^\Delta_{AB}$ of two detectors as a function of $\Delta z/\sigma$ with $\Omega_A\sigma=0.10$ and $L/\sigma=0.05$. In all plots, the pink and yellow dashed lines represent the steering quantities $S^{A\rightarrow B}$ and $S^{B\rightarrow A}$, respectively, in flat spacetime without any boundary.}
\label{Fig.5}
\end{figure}

\begin{figure}
\begin{minipage}[t]{0.5\linewidth}
\centering
\includegraphics[width=3.0in,height=5.2cm]{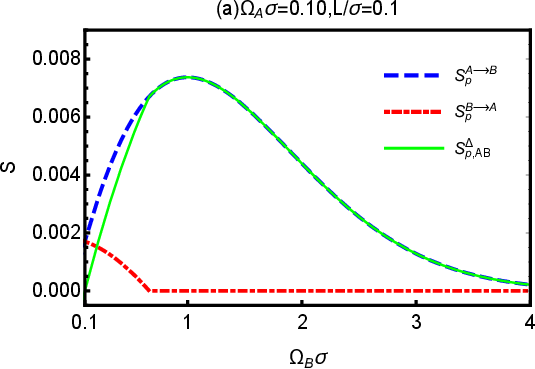}
\label{fig6a}
\end{minipage}%
\begin{minipage}[t]{0.5\linewidth}
\centering
\includegraphics[width=3.0in,height=5.2cm]{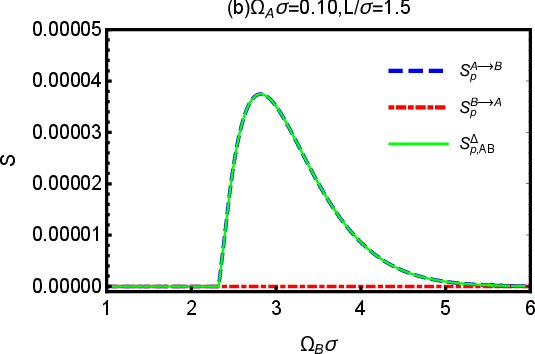}
\label{fig6b}
\end{minipage}%

\begin{minipage}[t]{0.5\linewidth}
\centering
\includegraphics[width=3.0in,height=5.2cm]{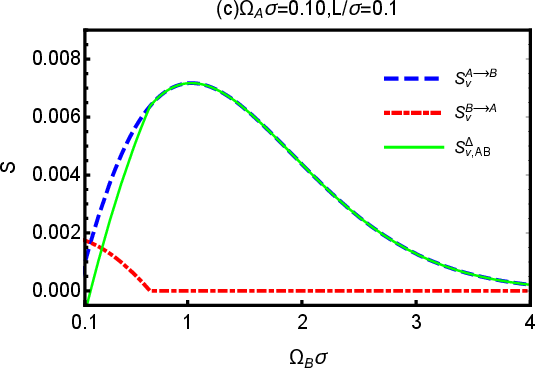}
\label{fig6c}
\end{minipage}%
\begin{minipage}[t]{0.5\linewidth}
\centering
\includegraphics[width=3.0in,height=5.2cm]{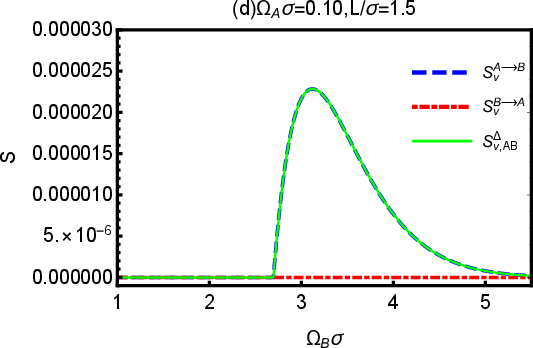}
\label{fig6d}
\end{minipage}%
\caption{The quantum steering $S^{A\rightarrow B}$, $S^{B\rightarrow A}$, and the steering asymmetry $S^\Delta_{AB}$ of two detectors as a function of the $\Omega_B\sigma$ with $\Omega_A\sigma=0.10$ and $\Delta z/\sigma$=1.00.}
\label{Fig.6}
\end{figure}

\begin{figure}
\begin{minipage}[t]{0.5\linewidth}
\centering
\includegraphics[width=3.0in,height=5.2cm]{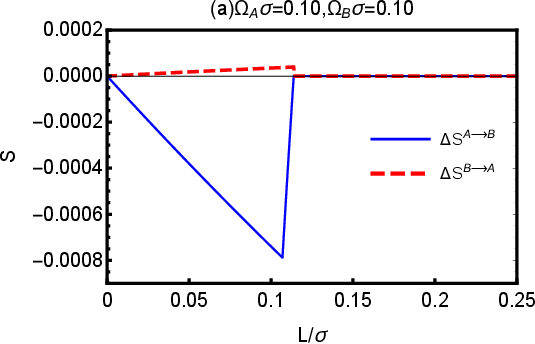}
\label{fig7a}
\end{minipage}%
\begin{minipage}[t]{0.5\linewidth}
\centering
\includegraphics[width=3.0in,height=5.2cm]{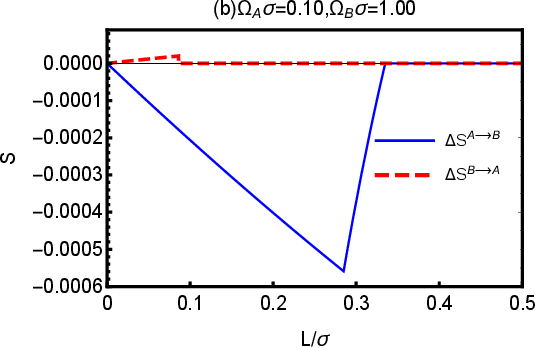}
\label{fig7b}
\end{minipage}%
\caption{The differences  $\Delta S^{A\rightarrow B}$ and $\Delta S^{B\rightarrow A}$ of two detectors as a function of $L/\sigma$ with $\Omega_A\sigma=0.10$ and $\Delta z/\sigma$=1.00.}
\label{Fig.7}
\end{figure}

We substitute Eqs.(\ref{S24}) and (\ref{S28}) into Eq.(\ref{S17}), and it becomes evident that $P_A$
takes the form of Eq.(\ref{S25}). Similarly, $P_B$ can be obtained by replacing $\Delta z$ in Eq.(\ref{S25}) with $\Delta z+L$. Using the same approach, we derive the auxiliary equations
$f (L)$ and $g (L)$ based on Eqs.(\ref{A1}) and (\ref{A2}). Furthermore, the correlation terms $C$ and $X$ can be calculated as
\begin{eqnarray}\label{S29}
C={\frac{\lambda^2}{4\sqrt{\pi}}}e^{-{\frac{(\Omega_B-\Omega _A)^2\sigma^2}{4}}}\big[f(L)-f(L+2\Delta z)\big],
\end{eqnarray}
\begin{eqnarray}\label{S30}
X=-{\frac{\lambda^2}{4\sqrt{\pi}}}e^{-{\frac{(\Omega_B+\Omega _A)^2\sigma^2}{4}}}\big[g(L)-g(L+2\Delta z)\big].
\end{eqnarray}
To compare the boundary-induced effects on steering in different configurations, we analyze how the orthogonal arrangement differs from the parallel case discussed earlier. The results are presented in Figs.\ref{Fig.4}-\ref{Fig.7}.

In Fig.\ref{Fig.4}, we plot the quantum steering $S^{A\rightarrow B}_V$, $S^{B\rightarrow A}_V$, and the steering asymmetry $S^\Delta_{V,{AB}}$ as a function of the detector separation $L/\sigma$ for different values of $\Omega_B\sigma$. For identical detectors, we can see that the steerability $S^{A\rightarrow B}_{V}$ of two detectors arranged orthogonally to the boundary is always smaller than the steerability $S^{B\rightarrow A}_{V}$ in Fig.\ref{Fig.4}(a), whereas for detectors arranged parallel to the boundary, the steerability $S^{A\rightarrow B}_{P}$ is always equal to the steerability $S^{B\rightarrow A}_{P}$. This difference arises because the detectors are asymmetrically positioned with respect to the boundary, leading to the steering asymmetry $S^\Delta_{V,{AB}}$. However, for non-identical detectors without a boundary, the steerability from Alice to Bob exceeds the steerability from Bob to Alice. Therefore, the conclusion of Fig.\ref{Fig.4}(a) is the opposite of the result in flat spacetime without a boundary. The dual effects of boundary and energy gap difference give the steering asymmetry $S^\Delta_{V,{AB}}$ richer properties in Fig.\ref{Fig.4}(b).
Specifically, for small detector separation $L/\sigma$, the steerability $S^{A\rightarrow B}_{V}$ is bigger than the steerability $S^{B\rightarrow A}_{V}$, indicating that the energy gap difference is the dominant factor. On the contrary, for larger detector separation $L/\sigma$, the steerability $S^{A\rightarrow B}_{V}$ is smaller than the steerability $S^{B\rightarrow A}_{V}$, implying that the boundary effect takes precedence. From  Fig.\ref{Fig.4}(c), we also see that,
for bigger  energy gap difference, the effect of energy gap difference  always plays an important role in the steerability. Comparing Fig.\ref{Fig.4}(a) and (c), we can observe that an increased energy gap enhances the harvest range of $S^{A\rightarrow B}_{V}$ while reducing the harvest range of $S^{B\rightarrow A}_{V}$.

In Fig.\ref{Fig.5}, we plot the $S^{A\rightarrow B}$, $S^{B\rightarrow A}$, and the steering asymmetry $S^\Delta_{{AB}}$ as a function of $\Delta z/\sigma$ for the two detectors in both parallel and orthogonal placements relative to the boundary, under different values of $\Omega_B\sigma$. The analysis of Fig.\ref{Fig.5} reveals a clear distinction in the steerability at the limit of $\Delta z/\sigma\rightarrow 0$ compared to without the boundary case, where the steerability is zero. This indicates a significant suppressive effect of the boundary on the detectors' steerability as it approaches the boundary. As $\Delta z/\sigma$ increases, this suppressive influence gradually transitions to a promotive effect, leading to a peak value where steering is maximally harvested. In the asymptotic limit, the steerability converges to the corresponding values in flat spacetime without a boundary. As shown clearly in Fig.\ref{Fig.5}(b) and (d), within the parameter region between the peak values of $S^{B \rightarrow A}$ and $S^{A \rightarrow B}$, increasing $\Delta z/\sigma$ leads to a decrease in $S^{B \rightarrow A}$ while $S^{A \rightarrow B}$ correspondingly increases. This indicates that the boundary  suppresses steering from $B$ to $A$ while enhancing steering from $A$ to $B$, thereby revealing the direction-dependent role of the boundary in modulating quantum steering.

To further explore how the harvesting of quantum steering by two non-identical detectors depends on their energy gap difference, we present the steering behavior as a function of this difference in Fig.\ref{Fig.6}. Specifically, Fig.\ref{Fig.6} shows the variation of $S^{A\rightarrow B}$, $S^{B\rightarrow A}$, and the steering asymmetry $S^\Delta_{{AB}}$ as a function of $\Omega_B\sigma$, for detectors aligned both parallel and orthogonal to the boundary.
Note that varying $\Omega_B\sigma$ is mathematically equivalent to tuning the relative energy gap difference $\Delta \Omega\sigma/\Omega_A$. From Fig.\ref{Fig.6}(a) and (c), we observe that for small values of $L/\sigma$, the asymmetry $S^\Delta_{AB}$ initially increases with $\Omega_B\sigma$, reaches a peak, and then drops to zero. This non-monotonic behavior indicates that non-identical detectors can enhance steering asymmetry harvested from the vacuum. Moreover, compared to identical detectors, non-identical detectors can harvest more $S^{A\rightarrow B}$, while the steerability $S^{B\rightarrow A}$ is suppressed. This suggests that the energy gap difference plays a dual role-enhancing steerability in one direction while inhibiting it in the other. In Fig.\ref{Fig.6}(b) and (d), corresponding to larger values of $L/\sigma$, the steerability $S^{A\rightarrow B}$ exhibits a ``sudden birth", followed by a growth to a peak, and then a ``sudden death" as $\Omega_B\sigma$ continues to increase. Notably, $S^{B\rightarrow A}$ remains zero throughout, indicating that in such configurations, non-identical detectors can only generate one-way steering from the vacuum field.

Finally, we investigate the influence of the boundary on the difference in quantum steering between two detector alignments. To quantify this effect, we define the differences as
\begin{eqnarray}\label{S31}
\Delta S^{A\rightarrow B}=S^{A\rightarrow B}_V-S^{A\rightarrow B}_P,
\end{eqnarray}
\begin{eqnarray}\label{S31}
\Delta S^{B\rightarrow A}=S^{B\rightarrow A}_V-S^{B\rightarrow A}_P,
\end{eqnarray}
where the subscripts $V$ and $P$ represent the orthogonal and parallel detector configurations with respect to the boundary, respectively.

Fig.\ref{Fig.7} shows how $\Delta S^{A\rightarrow B}$ and $\Delta S^{B\rightarrow A}$ vary as a function of $L/\sigma$. As seen from the plot, $\Delta S^{B\rightarrow A}$ first increases with $L/\sigma$, then gradually decreases to zero. This indicates that for small $L/\sigma$, the harvested $S^{B\rightarrow A}$ is greater when the detectors are orthogonally aligned, compared to the parallel configuration. In contrast, $\Delta S^{A\rightarrow B}$ follows a different trend: it initially decreases, then rises back toward zero as $L/\sigma$ increases. This implies that for smaller $L/\sigma$, the harvested $S^{A\rightarrow B}$ is larger in the parallel configuration than in the orthogonal one. These results suggest that the optimal arrangement of detectors for harvesting quantum steering from the vacuum depends on the direction of steering: to enhance steering from Bob to Alice, the detectors should be placed orthogonally to the boundary; whereas to maximize steering from Alice to Bob, they should be aligned parallel to the boundary.

\begin{figure}
\begin{minipage}[t]{0.5\linewidth}
\centering
\includegraphics[width=3.0in,height=5.2cm]{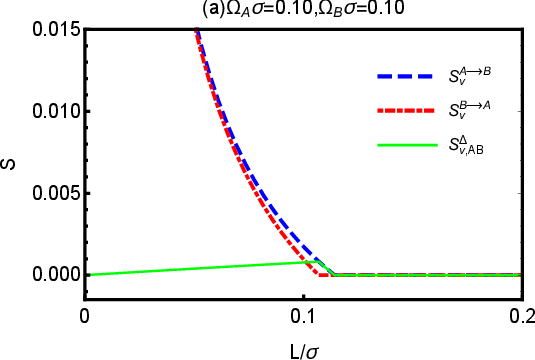}
\label{fig8a}
\end{minipage}%
\begin{minipage}[t]{0.5\linewidth}
\centering
\includegraphics[width=3.0in,height=5.2cm]{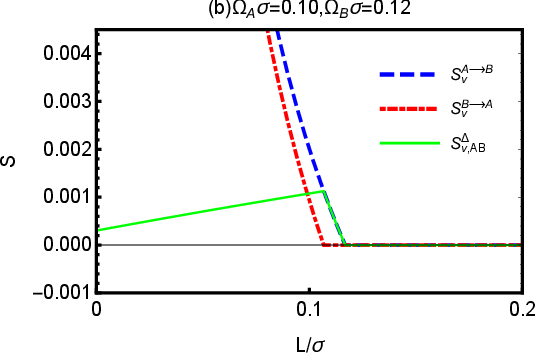}
\label{fig8b}
\end{minipage}%
\caption{Quantum steering $S^{A\rightarrow B}_V$, $S^{B\rightarrow A}_V$, and the steering asymmetry $S^\Delta_{V,{AB}}$ of two detectors as a function of $L/\sigma$ with $\Omega_A\sigma=0.10$ and $\Delta z/\sigma$=1.00.}
\label{Fig.8}
\end{figure}

Finally, to investigate the impact of swapping detectors $A$ and $B$ on quantum steering, we exchange their positions as depicted in Fig.\ref{Fig.3}. Fig.\ref{Fig.8} presents quantum steering $S^{A\rightarrow B}_V$, $S^{B\rightarrow A}_V$, and the steering asymmetry $S^\Delta_{V,{AB}}$ as a function of the detector separation $L/\sigma$ for different values of $\Omega_B\sigma$. From Fig.\ref{Fig.8}(a), we can see that for identical detectors aligned orthogonally to the boundary, the steerability $S^{A\rightarrow B}_V$ remains consistently greater than $S^{B\rightarrow A}_V$ as the separation $L/\sigma$ increases.
In addition, for non-identical detectors in free space (without a boundary),
the steerability from Alice to Bob remains stronger than the steerability from Bob
to Alice. Therefore, this trend persists even for non-identical detectors near the boundary, as confirmed in Fig.\ref{Fig.8}(b). However, in the configuration of Fig.\ref{Fig.4}, $S^{A\rightarrow B}_V$ can either surpass or fall below $S^{B\rightarrow A}_V$, which contrasts with the findings in Fig.\ref{Fig.8}.

\section{Conclusion}
In this work, we have conducted a comprehensive study of quantum steering harvesting between two static Unruh-DeWitt (UDW) detectors coupled to a vacuum massless scalar field in flat spacetime with a perfectly reflecting boundary. Our focus was to examine how both detector configuration (parallel vs orthogonal to the boundary) and energy gap differences influence the magnitude and directionality of the harvested steering.

Unlike entanglement, which is symmetric by definition and remains so even in the presence of a boundary, quantum steering inherently exhibits directional asymmetry. This directional sensitivity enables it to capture fine-grained spatial features of the quantum field. Our findings reveal several important physical insights: \textbf{(i) boundary-induced directional asymmetry:} the boundary breaks the spatial isotropy of the Minkowski vacuum, resulting in different field responses depending on detector positions. We observe that quantum steering is suppressed in one direction while being enhanced in the other as the detector-to-boundary distance increases. This effect is absent in entanglement harvesting and illustrates that steering is more sensitive to the environment; \textbf{(ii) configuration dependence:} for detectors aligned parallel to the boundary, steering is symmetric in the case of identical detectors and becomes asymmetric only due to energy gap differences. However, in the orthogonal configuration, the asymmetry arises even for identical detectors. This is due to the detectors' unequal proximity to the boundary, leading to asymmetric excitation probabilities and uneven access to the field's quantum correlations. This difference is physically meaningful: it shows that spatial asymmetry translates directly into steering asymmetry; \textbf{(iii)  sudden death and birth phenomena:} we report the ``sudden death" of steering as detector separation increases, indicating that the system transitions from two-way to one-way, and eventually no-way steering. Conversely, when increasing the energy gap of one detector, we observe a ``sudden birth" of one-way steering. These effects originate from the delicate balance between the local detector excitations and the nonlocal vacuum correlations, both of which are strongly modulated by the presence of the boundary; \textbf{(iv) energy gap difference as a control parameter:} the energy gap difference plays a dual role-enhancing steering in one direction while suppressing it in the other. This asymmetry suggests that quantum steering can be directionally controlled by tuning internal detector parameters, offering a practical handle for steering-based quantum information tasks; \textbf{(v) optimizing steering extraction:} our analysis of both configurations demonstrates that the optimal geometry for steering extraction is direction-dependent. If one aims to maximize steering from $A$ to $B$, a parallel configuration with suitable gap tuning is favorable. Conversely, for steering from $B$ to $A$, an orthogonal configuration near the boundary is more effective. These findings indicate that the boundary can be treated as a tunable resource to control and direct quantum correlations.

In conclusion, our work illustrates that quantum steering provides a more nuanced and physically sensitive probe of vacuum correlations than entanglement. It uncovers direction-dependent effects that are hidden in symmetric measures like entanglement or mutual information. These insights not only enrich the theoretical understanding of quantum field correlations in bounded spacetimes but also pave the way for practical steering-based protocols in relativistic quantum information, where detector geometry and internal parameters can be designed to selectively extract directional quantum resources.

\begin{acknowledgments}
This work is supported by the National Natural
Science Foundation of China (Grant Nos. 12205133) and  the Special Fund for Basic Scientific Research of Provincial Universities in Liaoning under grant NO. LS2024Q002.
\end{acknowledgments}


\end{document}